\documentclass[10pt,journal,cspaper,compsoc]{IEEEtran}

\usepackage[nocompress]{cite}
\usepackage{amssymb}
\usepackage{graphicx}
\usepackage{color}
\usepackage[hyphens]{url}
\usepackage{multirow}
\usepackage{rotating}
\usepackage{tikz}
\usepackage{hyperref}

\definecolor{linkblue}{RGB}{0,0,155}
\hypersetup{
   bookmarksnumbered,
   colorlinks={true},
   pdfstartview={FitH},
   citecolor={linkblue},
   linkcolor={linkblue},
   urlcolor={linkblue},
}

\newenvironment{ttquote}{\smallskip\list{}{\ttfamily\leftmargin=0mm\rightmargin=0cm}\small\item[]}{\endlist\smallskip}

\begin{document}

\title{Making Digital Artifacts on the Web\\Verifiable and Reliable}

\author{Tobias~Kuhn and Michel~Dumontier%
\IEEEcompsocitemizethanks{\IEEEcompsocthanksitem T. Kuhn is with the
Department of Humanities, Social and Political Sciences, ETH Zurich, Clausiusstrasse 37, 8092 Zurich, Switzerland. \texttt{tokuhn@ethz.ch}
\IEEEcompsocthanksitem M. Dumontier is with the Stanford Center for Biomedical Informatics Research, Stanford University, 1265 Welch Road, Stanford, CA 94305-5479, USA.\newline\texttt{michel.dumontier@stanford.edu}}%
\thanks{}}

\markboth{IEEE Transactions on Knowledge and Data Engineering
}{Kuhn and Dumontier: Trusty URIs}

\IEEEcompsoctitleabstractindextext{%
\begin{abstract}
The current Web has no general mechanisms to make digital artifacts --- such as datasets, code, texts, and images --- verifiable and permanent. For digital artifacts that are supposed to be immutable, there is moreover no commonly accepted method to enforce this immutability. These shortcomings have a serious negative impact on the ability to reproduce the results of processes that rely on Web resources, which in turn heavily impacts areas such as science where reproducibility is important. To solve this problem, we propose \emph{trusty URIs} containing cryptographic hash values.
We show how trusty URIs can be used for the verification of digital artifacts, in a manner that is independent of the serialization format in the case of structured data files such as nanopublications. We demonstrate how the contents of these files become immutable, including dependencies to external digital artifacts and thereby extending the range of verifiability to the entire reference tree. 
Our approach sticks to the core principles of the Web, namely openness and decentralized architecture, and is fully compatible with existing standards and protocols. Evaluation of our reference implementations shows that these design goals are indeed accomplished by our approach, and that it remains practical even for very large files.
\end{abstract}
\begin{IEEEkeywords}
Decentralized systems, data publishing, Semantic Web, Linked Data, Resource Description Framework, nanopublications
\end{IEEEkeywords}
}

\maketitle

\IEEEdisplaynotcompsoctitleabstractindextext

\IEEEpeerreviewmaketitle

\section{Introduction}

\PARstart{I}{n} many areas and in particular in science, reproducibility is important. Verifiable, immutable, and permanent digital artifacts are an important ingredient for making the results of automated processes reproducible, but the current Web offers no commonly accepted methods to ensure these properties. Endeavors such as the Semantic Web to publish complex knowledge in a machine-interpretable manner aggravate this problem, as automated algorithms operating on large amounts of data can be expected to be even more vulnerable than humans to manipulated or corrupted content. Without appropriate counter-measures, malicious actors can sabotage or trick such algorithms by adding just a few carefully manipulated items to large sets of input data. To solve this problem, we propose an approach to make items on the (Semantic) Web verifiable, immutable, and permanent.
This approach includes cryptographic hash values in Uniform Resource Identifiers (URIs) and adheres to the core principles of the Web, namely openness and decentralized architecture.
This article is an extended and revised version of a conference paper \cite{kuhn2014eswc}.

A cryptographic hash value (sometimes called \emph{cryptographic digest}) is a short random-looking sequence of bytes (or, equivalently, bits) that are calculated in a complicated yet perfectly predictable manner from a digital artifact such as a file. The same input always leads to exactly the same hash value, whereas just a minimally modified input returns a completely different value. While there is an infinity of possible inputs that lead to a specific given hash value, it is impossible in practice (for strong state-of-the-art hash algorithms) to reconstruct \emph{any} of the possible inputs just from the hash value. This means that if you are given some input and a matching hash value, you can be sure that the hash value was obtained from exactly that input. On this basis, our proposed approach boils down to the idea that references can be made completely unambiguous and verifiable if they contain a hash value of the referenced digital artifact. Our method does not apply to all URIs, of course, but only to those that are meant to represent a specific and immutable digital artifact.

Let us have a look at a concrete example: Nanopublications have been proposed as a new way of scientific publishing \cite{groth2010isu}. The underlying idea is that scientific results should be published not just as narrative articles but also in terms of minimal pieces of computer-interpretable results in a formal semantic notation (i.e. RDF). Nanopublications can cite other nanopublications via their URIs, thereby creating complex citation networks. Published nanopublications are supposed to be immutable, but there is currently no mechanism to enforce this. It is well-known that even artifacts that are supposed to be immutable tend to change over time, while often keeping the same URI reference. For approaches like nanopublications, however, it is important to specify exactly what version of what resource they are based on, and nobody should be given the opportunity to silently modify his or her already published contributions.

\begin{figure*}[t]
\begin{center}
\includegraphics[width=0.8\textwidth]{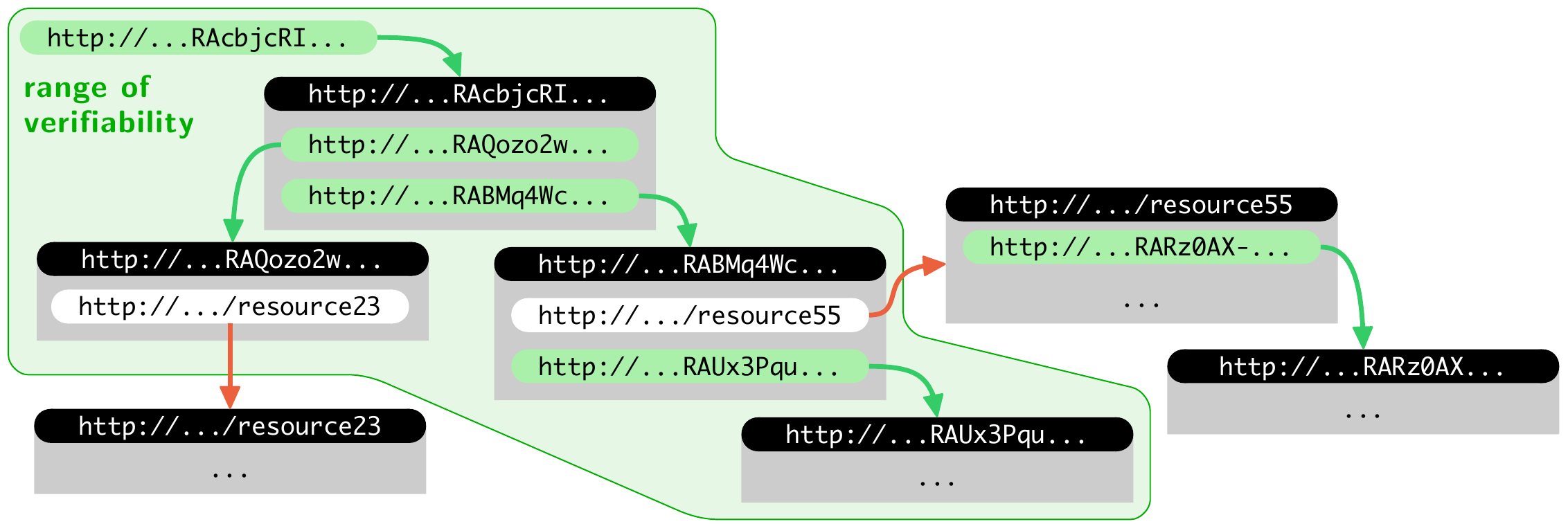}
\caption{Schematic illustration of the range of verifiability for the trusty URI on the top left. The green area shows its range of verifiability that covers all artifacts that can be reached by following trusty URI links (green arrows).}
\label{fig:verifiability}
\end{center}
\end{figure*}

With the approach outlined below, nanopublications can be identified with \emph{trusty URIs} that include cryptographic hash values calculated on the RDF content. For example, let us assume that you have a nanopublication with identifier $I_1$ that cites another nanopublication with identifier $I_2$. If you want to find the content of $I_2$, you can simply search for it on the Web, not worrying whether the source is trustworthy or not, and once you have found an artifact that claims to be $I_2$, you only have to check whether the hash value actually matches the content. If it does, you got the right nanopublication, and if not you have to keep searching (this can of course be automated). A trusty URI like $I_1$ does not only allow you to retrieve its nanopublication in a verifiable way, but in the next step also all nanopublications it cites (such as $I_2$) and all nanopublications they cite and so on. Any trusty URI in a way ``contains'' the complete backwards history that is reachable by following trusty URI links.
In this sense, the \emph{range of verifiability} is not just the resource itself, but the complete reference tree obtained by recursively following all contained trusty URIs. This is illustrated in Fig. \ref{fig:verifiability}.

Another important property of nanopublications is that they are self-con\-tained in the sense that they consist not only of the actual scientific assertions but also of their provenance information and meta-data. This means that nanopublications contain self-references in the form of their own identifying URIs. The calculation of a trusty URI must therefore allow for the resulting URI to be part of the digital artifact it is calculated on (this might sound impossible at first, but we show below how it can be achieved).
This leads us to the formulation of the following requirements:
\begin{enumerate}
\item To allow for verification of not only a given digital artifact but its entire reference tree, the hash should be part of the URI of the artifact.
\item To allow for the inclusion of meta-data, digital artifacts should be allowed to contain self-references (i.e. their own URIs).
\item The verification should be performed on a more abstract level than just the bytes of a file, with modules for different types of content. It should be possible to verify a digital artifact even if it is presented in a different format.
\item The complete approach should be decentralized and open: Everybody should be allowed to make verifiable URIs without a central authority.
\item The approach should be based on current established standards and be compatible with current tools and formats, so that it can be used right away.
\end{enumerate}
Though there are a number of related approaches, we are not aware of any general approach that complies with all these requirements. In particular, requirements 2 and 3 are not addressed by existing approaches.
The main benefits of artifacts with a trusty URI are that they are verifiable, immutable, and permanent. Let us briefly explain what we mean by these properties.

Trusty URI artifacts are \emph{verifiable} in the sense that a retrieved artifact for a given URI can be checked to contain the content the URI is supposed to represent. It can be detected if the artifact got corrupted or manipulated on the way, assuming that the trusty URI for the required artifact is known, e.g. because another artifact contains it as a link. (Of course, somebody can give you a manipulated artifact with a \emph{different} trusty URI.)

It directly follows that trusty URI artifacts are \emph{immutable}, as any change in the content also changes its URI, thereby making it a \emph{new} artifact. Again, you can of course change your artifact \emph{and} its URI and claim that it has always been like this. You can get away with that if the trusty URI has not yet been picked up by third parties, i.e. linked by other resources. Once this is the case, it cannot be changed anymore, because all these links will still point to the old trusty URI and everybody will notice that the new artifact is a different one.

Third, trusty URI artifacts are \emph{permanent} if we assume that there are search engines and Web archives crawling the artifacts on the Web and caching them. In this situation, any artifact that is available on the Web for a sufficiently long time will remain available forever. If an artifact is no longer available in its original location (e.g. the one its URI resolves to), one can still retrieve it from the cache of search engines, Web archives, or dedicated replication services. The trusty URI guarantees that it is the artifact you are looking for, even if the location of the cached artifact is not trustworthy or it was cached from an untrustworthy source.

\section{Background}

There are a number of related approaches based on cryptographic hash values.
The Git version control system ({\small\url{http://git-scm.com}}), for example, uses hash values to identify commits of distributed repositories.
For such a distributed repository, commits can happen asynchronously and anywhere (even when the respective site is offline). All sites need to be able to issue commit identifiers, yet these identifiers have to be unique. Hash values are a natural solution to this problem.
An important difference to our approach is that hash values (called \emph{checksums} in Git) are used to identify the respective artifacts (\emph{commits} in Git) only within a given repository and not on the Web scale. A second important difference is that the hash represents the byte content of files, whereas our approach allows for digital content at different levels of abstraction. On the technical side, Git uses the SHA-1 algorithm, which is no longer considered secure (which is not a serious problem for Git, because typically only trusted parties have write access to a repository).

The proposed standard for Named Information (ni) URIs \cite{keranen2013ietf} is another important related approach. It introduces a new URI protocol \texttt{ni} to refer to digital artifacts with hash values in a uniform way. These are two examples of ni-URIs:
\begin{ttquote}
ni:///sha-256;UyaQV-Ev4rdLoHyJJWCi11OHfrYv9E1a\\
\verb|  |GQAlMO2X\_-Q\\
ni://example.org/sha-256;5AbXdpz5DcaYXCh9l3eI9\\
\verb|  |ruBosiL5XDU3rxBbBaUO70
\end{ttquote}
The ni-URI approach allows for different hash algorithms, such as SHA-256 (which is, in contrast to SHA-1, considered secure) and optional specification of an authority, such as \texttt{example.org}, where the artifact can be found. It misses, however, some of the features of our list of requirements. As with Git, ni-URIs do not define how digital artifacts can be represented at a more abstract level than their sequence of bytes, and self-references are not supported. Furthermore, current browsers do not recognize the \texttt{ni} protocol, and administrator access to a server is needed to make these URIs resolvable. The latter two points are not a real problem in the long run, but they might hinder the adoption of the standard in the first place. The approach presented in this paper is complementary and compatible. We propose trusty URIs, which can be mapped to ni-URIs but are more flexible and provide additional features.

There are a number of existing approaches to include hash values in URIs for verifiability purposes, e.g. for legal documents \cite{hoekstra2011iswc}. The downside of such custom-made solutions is that custom-made software is required to generate, resolve, and check the hash references.
Here, we propose a general approach that could replace such specific ones, thereby establishing interoperability of systems and standard infrastructure for creating, resolving, and checking hash references.
Standards have been proposed for the verification of quantitative datasets \cite{altman2007dlib} and XML documents \cite{bartel2008xmldsig}, but they are not general enough to cover content such as RDF graphs (at least not in a convenient way) and they keep the hash value separate from the URI reference, which means that the range of verifiability does not directly extend to referenced artifacts.

To calculate hash values on content that is more abstract than just a fixed sequence of bytes, common approaches require the normalization (also called \emph{canonicalization}) of the respective data structures such as RDF graphs. In the general case, RDF graph normalization is known to be a very hard problem, possibly unsolvable in polynomial time \cite{carroll2003iswc}. General algorithms exist \cite{hofig2014ursw} but they are very inefficient in the general case, with a runtime complexity of $O(n^n)$.
This stems from the difficulty of handling blank nodes, i.e. identifiers that are only unique in a local scope and can be locally renamed without effects on semantics. Without blank nodes, normalization boils down to sorting of RDF triples, which can be performed in $O(n \log n)$.
The need for sorting can even be eliminated by using incremental cryptography \cite{bellare1994crypto}, which allows for calculating digests for RDF graphs without blank nodes in linear time \cite{sayers2004rdfdigets}. Such incremental approaches, however, are not as well-studied as mainstream cryptography methods, and open the possibility of new kinds of attacks \cite{phan2006cs}.
Efficient normalization algorithms that support blank nodes put restrictions on the graph structure and require additional (semantically neutral) triples to be added to some graphs before they can be processed \cite{carroll2003iswc,sayers2004rdfdigets}.

The general need for persistent identifiers for scientific artifacts is widely acknowledged and tackled by a number of approaches \cite{vandesompel2014digitalcuration}.
Similar methods to the ones presented in this paper, i.e. calculating hash values in a format-independent manner, have been proposed to track the provenance of data sets \cite{mccusker2012ieee}. This has been used to define a conceptualization of multi-level identities for digital works based on cryptographic digests and formal semantics, covering different conceptual levels from single HTTP transactions to high-level content identifiers \cite{mccusker2012ipaw}.

With respect to the general goal of making scientific results more reproducible, there are notable efforts in diverse areas such as bioinformatics \cite{gentleman2005sagmb}, computer science \cite{peng2011science}, and psychology \cite{open2012pps}. Research Objects ({\small\url{http://www.researchobject.org}}) \cite{bechhofer2010future} are a proposal to bundle papers with their datasets, code, workflows, logs, and other relevant metadata. Such Research Objects are self-contained and immutable packages that are sharable and cite-able, and should make the respective research reusable and reproducible. For this and similar approaches, trusty URIs could be used in the future to make such bundles and other kinds of digital artifacts verifiable and to enforce their immutability.

\section{Approach}

We propose here a modular approach, where different modules handle different kinds of content on different conceptual levels of abstraction, from byte level to high-level formalisms. Besides that, the most interesting features of our approach are self-references, the handling of blank nodes, and the mapping to ni-URIs.

\subsubsection*{General Structure}

Trusty URIs end with a hash value in Base64 notation (a specific alphanumeric encoding scheme) preceded by a module identifier.
This is an example:
\begin{ttquote}
http://example.org/r1.RA5AbXdpz5DcaYXCh9l3eI9r\\
\verb|  |uBosiL5XDU3rxBbBaUO70
\end{ttquote}
Everything that comes after \texttt{r1.} is the part that is specific to trusty URIs, which we call \emph{artifact code}. Its first two characters \texttt{RA} identify the module specifying its type (first character) and version (second character). The remaining 43 characters represent the actual hash value. The precise specification of trusty URIs is given below in Section \ref{sec:spec}.
Importantly, our approach entails a certain shift of authority: Once a trusty URI is established, its artifact code defines what object it refers to, and the issuing authority has no longer the power to change its meaning.

\subsubsection*{Self-References}

\newcommand{\hashplaceholder}{{\setlength{\fboxsep}{0.5mm}\hspace{0.2mm}\fbox{\normalfont\textsc{c}}\hspace{0.2mm}}}

To support self-references, i.e. resources that contain their own trusty URI, the generation process involves not just to compute the hash from a given artifact but to actually transform the artifact into a new version that contains the newly generated trusty URI.
For example, a resource like \texttt{\small{}http://example.org/r2} might have the following RDF content with a self-reference:
\begin{ttquote}
<http://example.org/r2> dct:description\\
\verb|  |"something" .
\end{ttquote}
To transform such a resource, we first define the structure of the new trusty URI by adding a placeholder {\hashplaceholder} where the artifact code should eventually appear. In the given example, the content would then look like this:
\begin{ttquote}
<http://example.org/r2.{\hashplaceholder}> dct:description\\
\verb|  |"something" .
\end{ttquote}
Note that it is necessary to add a non-Base64 character (in this case a dot ``\texttt{.}'') as a delimiter in front of {\hashplaceholder} if it would otherwise be preceded by a Base64 character. On such content, we can calculate a hash value by interpreting the placeholder {\hashplaceholder} as a blank space (the result is unambiguous as URIs are not allowed to otherwise contain blank spaces). Then we can replace the placeholder by the calculated artifact code and we end up with a trusty URI like this:
\begin{ttquote}
http://example.org/r2.RATf-GlZsJa1v\_EG0-yl5jwc\\
\verb|  |GNPF5zRbhDifBLeG4Q57c
\end{ttquote}
For strong hashing algorithms, it is impossible in practice that this calculated sequence of bytes was already part of the original content before the transformation. This entails that the replacing of the placeholder is reversible.

This reversibility is needed once an existing trusty URI resource containing self-references should be verified. We can revert the transformation described above by replacing all occurrences of the artifact code with a blank space, and then calculate the hash in the same way as when a resource is transformed. The content is successfully verified if and only if the resulting hash matches the one from the trusty URI.

\subsubsection*{Blank Nodes}

The support for self-references requires us to transform the preliminary content of a trusty URI artifact into its final version, and we can make use of this transformation to also solve the problem of blank nodes in RDF.
A blank node is basically an identifier that is only used in a local scope and for which we do not care to specify a concrete URI.
Our approach is to eliminate blank nodes during the transformation process by converting them into URIs. Blank nodes can be seen as existentially quantified variables, which we can turn into constants by Skolemization, i.e. by introducing URIs that have not been used anywhere before.
Using the trusty URI with a suffix enumerating the blank nodes, we can create such URIs guaranteed to have never been used before (the artifact code being just a placeholder at first, as above):
\begin{ttquote}
http://example.org/r3.RACjKTA5dl23ed7JIpgPmS0E\\
\verb|  |0dcU-XmWIBnGn6Iyk8B-U\#\_1\\
http://example.org/r3.RACjKTA5dl23ed7JIpgPmS0E\\
\verb|  |0dcU-XmWIBnGn6Iyk8B-U\#\_2
\end{ttquote}
This approach solves the problem of blank nodes for normalization, is completely general (i.e. works on any possible input graph), fully respects RDF semantics, and does not require auxiliary triples to be added.

\subsubsection*{ni-URIs}

Our approach is compatible with ni-URIs (see above), and all trusty URIs can be transformed into ni-URIs, with or without explicitly specifying an authority:
\begin{ttquote}
ni:///sha-256;5AbXdpz5DcaYXCh9l3eI9ruBosiL5XDU\\
\verb|  |3rxBbBaUO70\\
ni://example.org/sha-256;5AbXdpz5DcaYXCh9l3eI9\\
\verb|  |ruBosiL5XDU3rxBbBaUO70
\end{ttquote}
The fact that the module identifier is lost does not affect the uniqueness of the hash, but to verify a resource all available modules have to be tried in the worst case. To avoid this, we propose to use an optional argument called \texttt{module}:
\begin{ttquote}
ni:///sha-256;5AbXdpz5DcaYXCh9l3eI9ruBosiL5XDU\\
\verb|  |3rxBbBaUO70?module=RA
\end{ttquote}

\subsubsection*{Modules}

There are currently two module types available: F for representing byte-level file content and R for RDF graphs. For type F, the only version available as of now is A. For type R, there are two version A and B. This leads to the module identifiers \texttt{FA}, \texttt{RA}, and \texttt{RB}.

The difference between the two versions of module type R is that \texttt{RA} supports multiple graphs whereas \texttt{RB} requires a single RDF graph. While the former is more general, the latter provides more information about what the URI stands for. If a trusty URI of type \texttt{RB} is found in the graph position within a file or a triple store, it is immediately clear what it is supposed to stand for: the triples of the respective graph. For \texttt{RA} one only knows that it stands for the triples of a set of RDF graphs but additional information is needed (e.g. provided by the nanopublication definition) to find out \emph{which} graphs.

A certain module can be defined in a way that makes artifacts and their URIs \emph{transferable} to another module, in the sense that the module identifier can generally be switched to the second module without changing the hash or breaking the verification of the artifact. For example, module \texttt{RB} is transferable to module \texttt{RA}, but not vice versa.

Note that for an RDF document, either of the module types F and R could be used. The right choice depends on what the URI should identify. If it should identify a \emph{file} in a particular format and containing a fixed number of bytes, then F should be used. If it should, however, identify \emph{RDF content} independently of its serialization in a particular file, then R should be used.
For modules such as \texttt{RA} and \texttt{RB} that operate not just on the byte level, \emph{content negotiation} can be used to return the same content in different formats (depending on the requested content type in the HTTP request) when a trusty URI is accessed.

Even though we focus on RDF in this paper, the approach and architecture of trusty URIs are general and we plan to provide modules for additional kinds of content in the future. This could include tabular or matrix content (e.g. CSV or Excel files), content with tree structure (e.g. XML), hypertext (e.g. HTML or Markdown), bitmaps (e.g. PNG or JPEG), and vector graphics (e.g. SVG). New modules might also become necessary if the used hash algorithms should become vulnerable to attacks in the future.

\section{Specification}
\label{sec:spec}

This section contains version 1 of the specification of the \emph{trusty URI}
approach.\footnote{\url{http://trustyuri.net/spec/v1.FADQoZWcYugekAb4jW-Zm3_5Cd9tmkkYEV0bxK2fLSKao.md}}
The previous version was version 0, which was preliminary and
not stable.\footnote{\url{http://trustyuri.net/spec/v0.FA4BwXfTl2X-ABWKUF2k0T044yS2-KmO_R0zBftSsc96k.md}}

Version 1 described here is \emph{not} backward compatible with version 0 for module \texttt{RA} (the
handling of plain literals according to RDF 1.1 breaks compatibility). Future
versions, however, are supposed to be 100\% backward compatible with this
version. This means that when the extension of modules cannot be done in a
backward compatible manner in the future, new modules will have to be defined
and old ones will have to be deprecated.

\begin{figure*}
\begin{center}
\includegraphics[width=0.95\textwidth]{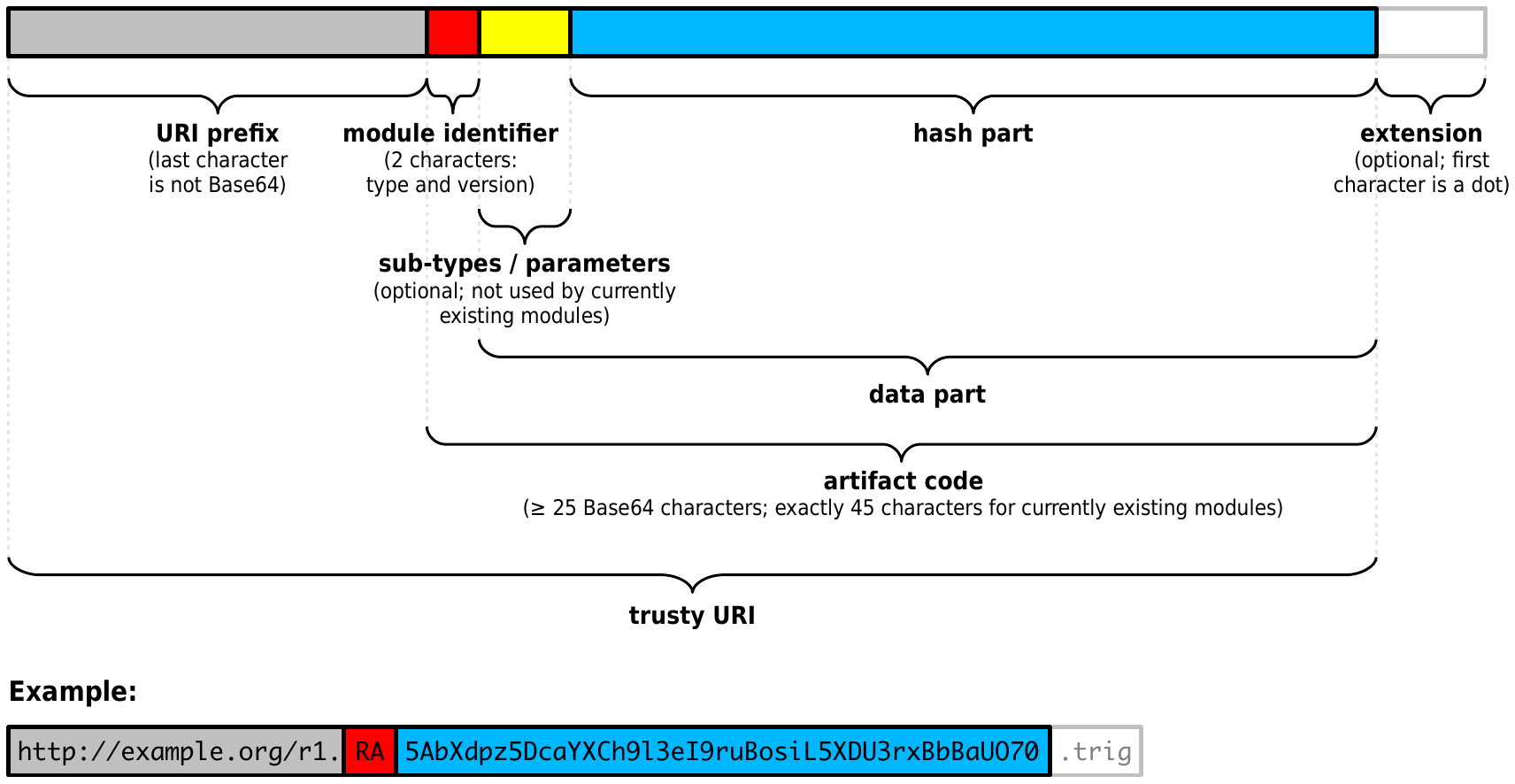}
\caption{Schematic representation of the general structure of trusty URIs.}
\label{fig:scheme}
\end{center}
\end{figure*}

Fig. \ref{fig:scheme} shows a schematic representation of the general structure of trusty URIs.

\subsection{Basics}

Hash values of trusty URIs are encoded in Base64 notation with some common
modifications for making it safe to use them in URIs and file names:
\begin{quote}
\textbf{Definition 1.}
Every character that is a standard ASCII letter (\texttt{A-Z} or \texttt{a-z}), a digit
(\texttt{0-9}), a hyphen (\texttt{-}), or an underscore (\texttt{\_}) is called a \emph{Base64
character}, representing in this order the numbers from 0 to 63. There are no
other Base64 characters.
\end{quote}
Trusty URIs have the following structure:
\begin{quote}
\textbf{Definition 2.}
Every trusty URI ends with at least 25 Base64 characters. The sequence of
characters following the last non-Base64 character is called the \emph{artifact
code}. The first two characters of the artifact code are called the \emph{module
identifier}. The sequence of characters following the module identifier
is called \emph{data part}, which is identical to or contains a \emph{hash part}.
\end{quote}
The current modules only generate URIs with exactly 45 trailing Base64
characters, but the definition is kept open for future modules.

The first character of the module identifier specifies the type of the content
and therefore the type of the module; the second character is a version identifier
of the module. The main content of the data part is the hash value, but
it can also contain other information such as parameters and sub-types. Its
concrete structure depends on the module.

As everybody who has access to the respective domain is free to define and use
URIs at will, we can only be sure that a certain URI is a trusty URI once we
have found and verified a content that matches the hash. For that reason, the concept of a \emph{potential trusty URI} needs to be introduced:
\begin{quote}
\textbf{Definition 3.}
Every URI that could be a trusty URI according to the restrictions of
Definition 2 with a module identifier matching a defined module and
with a data part that is consistent with the structural restrictions of the given
module (in particular with respect to its length) is called a \emph{potential
trusty URI}.
\end{quote}
With these ingredients, trusty URIs can be verified:
\begin{quote}
\textbf{Definition 4.}
Given a potential trusty URI and a digital artifact, if the identifier part
refers to an module that returns a hash value for the digital artifact
that is identical to the one encoded in the hash part, then the potential
trusty URI is a \emph{verified trusty URI} and the digital artifact is its
\emph{verified content}.
\end{quote}

For convenience reasons, we can append a file extension like \texttt{.txt} or \texttt{.nq}
to trusty URIs. The resulting URIs are technically no trusty URIs anymore, but
it is easy to strip the extension and get the respective trusty URIs.

As the hash values are located in the final part of the URIs, it is straightforward to
also use them in file names and to deal with them in a local file system without
worrying about the first part of the URI. For example, the name of such a file could therefore be:
\begin{ttquote}
r1.RA5AbXdpz5DcaYXCh9l3eI9ruBosiL5XDU3rxBbBaUO\\
\verb|  |70.nq
\end{ttquote}
Such files are called \emph{trusty files}.

\subsection{Modules}

There are currently three modules available: \texttt{FA}, \texttt{RA}, and \texttt{RB}.

\subsubsection*{Module FA}

Version A of module type F, i.e. module \texttt{FA}, works on the byte content of
files.
A hash value is calculated using SHA-256 \cite{nist2012shs} on the content of the file in byte
representation. The file name and other metadata are not considered. Two
zero-bits are appended to the resulting hash value, and then transformed to
Base64 notation as defined above. The resulting 43 characters make up the data
part of the trusty URI.

Empty files, for example, get the following URI suffix:%
\begin{ttquote}
FA47DEQpj8HBSa-\_TImW-5JCeuQeRkm5NMpJWZG3hSuFU
\end{ttquote}
When adding such a suffix to a URI, it has to be made sure that it
is preceded by a non-Base64 character, such as a dot (\texttt{.}), a slash (\texttt{/}), or a
hash sign (\verb|#|). This applies to all modules.

\subsubsection*{Module RA}

Version A of module type R, i.e. module \texttt{RA}, works on RDF content, possibly
covering multiple named graphs, relying on RDF version 1.1 \cite{cyganiak2014rdf}.

This module allows for self-references, i.e. the trusty URI itself may appear
in the RDF data it represents. URIs consisting of the given trusty URI and a
suffix are also supported, such as:
\begin{ttquote}
http://example.org/r2.RA5AbXdpz5DcaYXCh9l3eI9r\\
\verb|  |uBosiL5XDU3rxBbBaUO70\#Part1
http://example.org/r2.RA5AbXdpz5DcaYXCh9l3eI9r\\
\verb|  |uBosiL5XDU3rxBbBaUO70\#Part2
\end{ttquote}
Blank nodes are not supported and have to be skolemized in the same way when a trusty URI is
produced.
It is furthermore assumed that the data is a set of named RDF graphs. RDF triples without a
named graph are considered to belong to a special named graph represented with
the empty string.

To check whether a given artifact code correctly represents a given set of
named graphs, the triples and graphs have to be sorted first. Because the
trusty URI can appear in the RDF data it represents, all occurrences of the given artifact code in
the URIs have to be replaced by a blank character in a preprocessing step. To
determine the order of any two triples, the first applicable rule of the
following list is applied:

\begin{enumerate}
\item If their graph URIs differ, the triple with the lexicographically smaller
   preprocessed graph URI is first.
\item If their subject URIs differ, the triple with the lexicographically smaller
   preprocessed subject URI is first.
\item If their predicate URIs differ, the triple with the lexicographically
   smaller preprocessed predicate URI is first.
\item If one has a literal as object and the other has a non-literal, the triple
   with the non-literal as object is first.
\item If both have a URI as object, the triple with the lexicographically smaller
   preprocessed object URI is first.
\item If the literal labels of the objects differ, the triple with the
   lexicographically smaller literal label is first.
\item If one of the object literals has a datatype identifier and the other does
   not, the triple without a datatype identifier is first.
\item If one of the object literals has a language identifier and the other does
   not, the triple without a language identifier is first.
\item The triple with the lexicographically smaller datatype or language
   identifier is first.
\end{enumerate}
The lexicographic order is defined on strings of Unicode characters. If two
strings have different characters at at least one position, the string with the
smaller integer value at the first differing position is first. Otherwise, the
shorter string is first.

After the triples have been sorted, a sequence of Unicode characters $s$ is
built. For each triple, the serialization of its graph, its subject, its
predicate, and its object are added to the end of $s$, in this order and with a
newline character at the end of each of the four. The serialization of graph,
subject, and predicate identifiers is simply their preprocessed URI string.
Objects that consist of a URI are treated the same way. Literals without a
language tag are serialized as a circumflex character (\verb|^|) followed by the
datatype URI (which, according to RDF 1.1, equals
\verb|http://www.w3.org/2001/XMLSchema#string| if not explicitly specified), a blank
space, and the escaped literal string. Literal strings are escaped by replacing
\verb|\| by \verb|\\| and newline characters by 
\verb|\n|. Literals with a language tag are
serialized as an at-sign \verb|@| followed by the canonicalized language string, a
blank space, and the escaped literal string.

The actual computation of the hash data is identical to Module F: a SHA-256
hash is generated for $s$ in UTF-8 encoding, two zero-bits are appended, and
the result is transformed to Base64 notation.

\subsubsection*{Module RB}

Version B of module type R, i.e. module \texttt{RB}, is a slight variation of module \texttt{RA}. While module \texttt{RA} can represent any number of RDF graphs, a trusty URI using module \texttt{RB} always represents just one graph. The calculation of the hash is identical to the procedure described above, with the only restriction being that all triples have the trusty URI of the given resource as their graph URI.

\section{Implementation}

There are currently three trusty URI implementations in the form of code libraries in Java, Perl, and Python.\footnote{\url{https://github.com/trustyuri/trustyuri-java},\\\url{https://github.com/trustyuri/trustyuri-perl},\\\url{https://github.com/trustyuri/trustyuri-python}}
The Java implementation uses the \emph{Sesame} library \cite{broekstra2002iswc} for RDF processing, the Perl implementation makes use of the \emph{Trine} package, and the Python implementation uses \emph{RDFLib}.\footnote{\url{http://search.cpan.org/dist/RDF-Trine/},\\\url{https://github.com/RDFLib/rdflib}}

These implementations provide a number of common functions for the different modules and formats. Currently, the following functions are available:
\begin{itemize}
\item \textbf{CheckFile} takes a file and validates its hash by applying the respective module.
\item \textbf{ProcessFile} takes a file, calculates its hash using module \texttt{FA}, and renames it to make it a trusty file.
\item \textbf{TransformRdf} takes an RDF file and a base URI, and transforms the file into a trusty file using a module of type R.
\item \textbf{TransformLargeRdf} is the same as above but using temporary files instead of loading the entire content into memory.
\item \textbf{CheckLargeRdf} checks an RDF file using module \texttt{RA} without loading the whole content into memory but using temporary files instead.
\item \textbf{CheckSortedRdf} checks an RDF file assuming that it is already sorted (and raises an error otherwise). The current implementations generate such sorted files by default, but this is not required by the specification.
\item \textbf{RunBatch} reads commands (any of the above) from a file and executes them one after the other.
\end{itemize}
These libraries for the different programming languages are still work in progress: All these functions are currently supported by the Java implementation, but the other libraries still lack some of them. Nanopublication-specific functions now are provided by the official Java library for nanopublications\footnote{\url{https://github.com/Nanopublication/nanopub-java}} using the trusty URI library.

\begin{figure*}
\begin{center}
\includegraphics[width=0.9\textwidth]{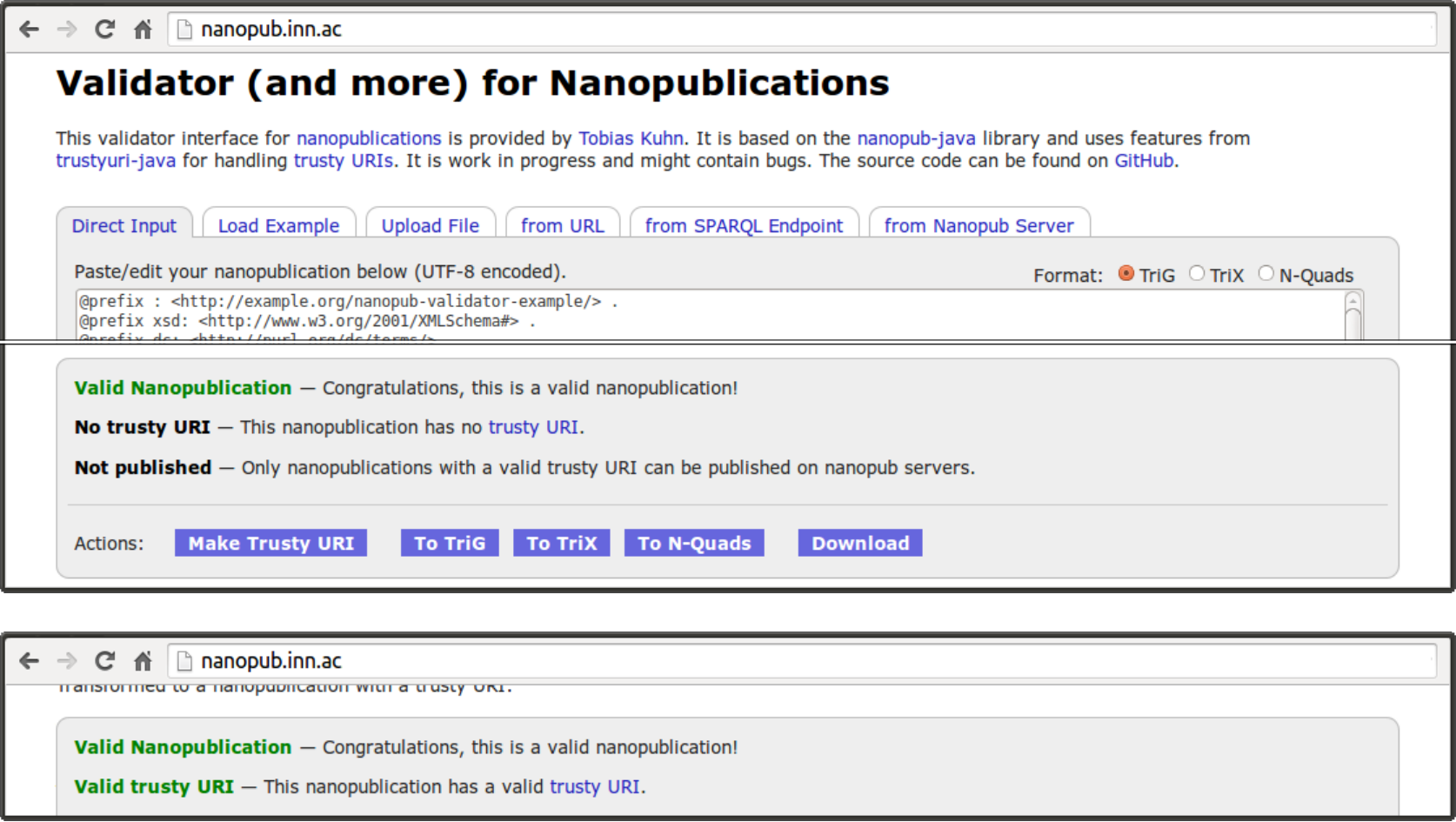}
\caption{The nanopublication validator interface integrates features of trusty URIs. Nanopublications can be loaded from different sources, users can generate trusty URIs for them (top), and nanopublications with trusty URIs are automatically verified (bottom).}
\label{fig:validator}
\end{center}
\end{figure*}

The trusty URI features provided by the presented libraries are also made available via a validation interface for nanopublications.\footnote{\url{http://nanopub.inn.ac/}} This interface, which is shown in Fig. \ref{fig:validator}, offers in fact much more than just validation (including transformation into different formats and publication to nanopublication servers). Users can load nanopublications in different ways, including retrieval from URLs or SPARQL endpoints, and then trusty URIs can be generated for them directly via the Web interface. Nanopublications that already have a trusty URI are automatically verified and users are informed about whether the verification was successful or not.

\section{Evaluation}

Below we present some experiments on the trusty URI concept and its implementations, based on two collections of RDF files.
\begin{table*}[tp]
\begin{center}
\caption{Performance and results of the different implementations for checking trusty URI nanopublications in normal mode (top) and batch mode (bottom) on valid and corrupted files.}
\label{tab:checkrdf}
{\scriptsize\sffamily
\vspace{3mm}
\renewcommand{\arraystretch}{1.75}
\textsc{Normal Mode}\smallskip\\
\begin{tabular}{r|lr|rrrrc|rrr}
\hline
 & \multicolumn{2}{c|}{method} & \multicolumn{5}{c|}{time in seconds} & \multicolumn{3}{c}{result} \\
 & impl. & format & mean & stdev & min & max & histogram & valid & invalid & error \\
\hline
\multirow{7}{*}{\begin{sideways}\phantom{p}valid files\phantom{p}\end{sideways}}
 & Java & N-Quads & 0.5229 & 0.0591 & 0.3750 & 5.5420& \multirow{7}{*}{\includegraphics[trim=5mm 13mm 1mm 11mm, clip=true, scale=0.49]{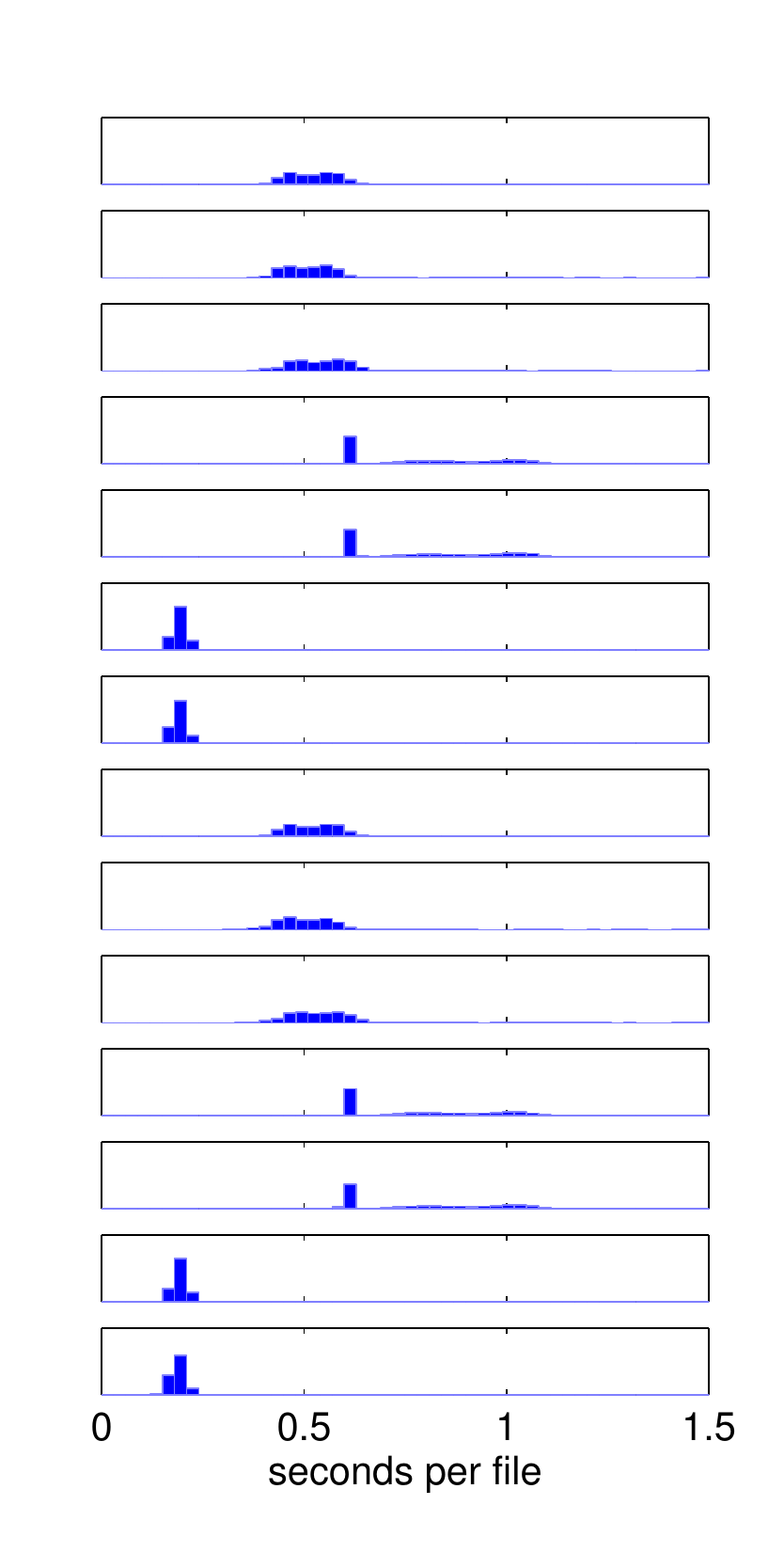}} & 100\% & 0\% & 0\% \\
 & Java & TriG & 0.5113 & 0.0569 & 0.3650 & 5.5340 & & 100\% & 0\% & 0\% \\
 & Java & TriX & 0.5383 & 0.0648 & 0.3900 & 5.5240 & & 100\% & 0\% & 0\% \\
 & Perl & N-Quads & 0.7843 & 0.1713 & 0.5990 & 5.7960 & & 100\% & 0\% & 0\% \\
 & Perl & TriG & 0.7901 & 0.1734 & 0.6030 & 5.7840 & & 100\% & 0\% & 0\% \\
 & Python & N-Quads & 0.1935 & 0.0164 & 0.1150 & 0.3050 & & 100\% & 0\% & 0\% \\
 & Python & TriX & 0.1912 & 0.0162 & 0.1190 & 0.3460 & & 100\% & 0\% & 0\% \\
\hline
\multirow{7}{*}{\begin{sideways}corrupted files\end{sideways}}
 & Java & N-Quads & 0.5227 & 0.0591 & 0.3450 & 5.5420 & & 0\% & 99.72\% & 0.28\% \\
 & Java & TriG & 0.5003 & 0.0621 & 0.3200 & 5.4250 & & 0\% & 83.37\% & 16.63\% \\
 & Java & TriX & 0.5322 & 0.0655 & 0.3360 & 5.5230 & & 0.83\% & 84.15\% & 15.03\% \\
 & Perl & N-Quads & 0.7842 & 0.1712 & 0.6000 & 5.8880 & & 0\% & 100\% & 0\% \\
 & Perl & TriG & 0.7872 & 0.1727 & 0.5700 & 5.8230 & & 0\% & 84.49\% & 15.51\% \\
 & Python & N-Quads & 0.1934 & 0.0165 & 0.1200 & 0.3080 & & 0\% & 100\% & 0\% \\
 & Python & TriX & 0.1884 & 0.0176 & 0.1070 & 0.2760 & & 0.12\% & 84.46\% & 15.42\% \\
\end{tabular}
\vspace{6mm}\\
\textsc{Batch Mode}\smallskip\\
\begin{tabular}{r|lr|rrrrc|rrr}
\hline
 & \multicolumn{2}{c|}{method} & \multicolumn{5}{c|}{time in seconds} & \multicolumn{3}{c}{result} \\
 & impl. & format & mean & stdev & min & max & histogram & valid & invalid & error \\
\hline
\multirow{7}{*}{\begin{sideways}\phantom{p}valid files\phantom{p}\end{sideways}}
 & Java & N-Quads & 0.0019 & 0.0062 & 0.0013 & 1.7202 & \multirow{7}{*}{\includegraphics[trim=5mm 13mm 1mm 11mm, clip=true, scale=0.49]{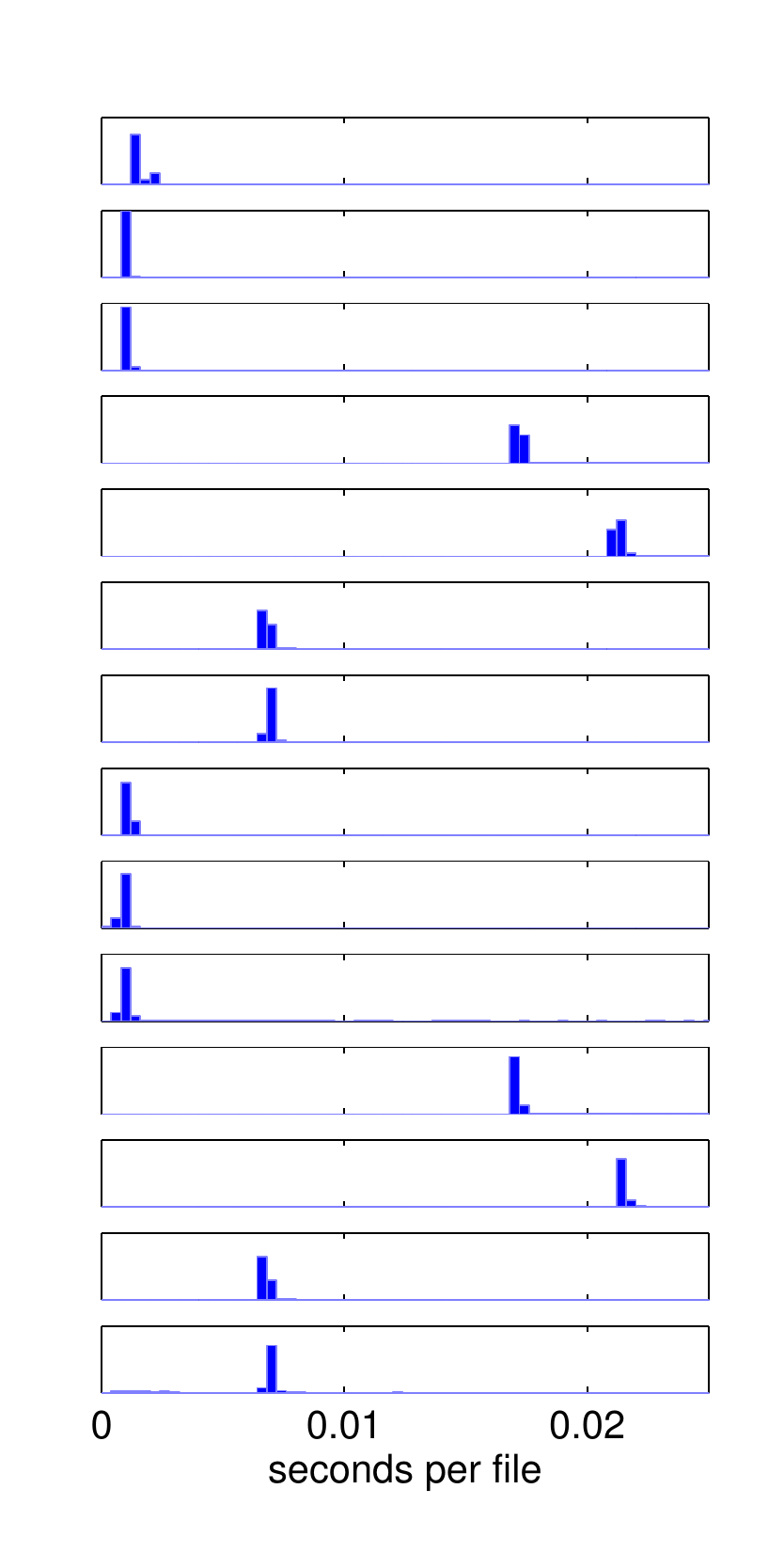}} & 100\% & 0\% & 0\% \\
 & Java & TriG & 0.0009 & 0.0050 & 0.0008 & 1.7412 & & 100\% & 0\% & 0\% \\
 & Java & TriX & 0.0011 & 0.0050 & 0.0009 & 1.5656 & & 100\% & 0\% & 0\% \\
 & Perl & N-Quads & 0.0172 & 0.0006 & 0.0171 & 0.0679 & & 100\% & 0\% & 0\% \\
 & Perl & TriG & 0.0214 & 0.0016 & 0.0211 & 0.0872 & & 100\% & 0\% & 0\% \\
 & Python & N-Quads & 0.0070 & 0.0011 & 0.0065 & 0.0644 & & 100\% & 0\% & 0\% \\
 & Python & TriX & 0.0070 & 0.0009 & 0.0066 & 0.0578 & & 100\% & 0\% & 0\% \\
\hline
\multirow{7}{*}{\begin{sideways}corrupted files\end{sideways}}
 & Java & N-Quads & 0.0012 & 0.0062 & 0.0006 & 1.6559 & & 0\% & 99.72\% & 0.28\% \\
 & Java & TriG & 0.0010 & 0.0049 & 0.0003 & 1.6335 & & 0\% & 83.37\% & 16.63\% \\
 & Java & TriX & 0.0011 & 0.0044 & 0.0005 & 1.3451 & & 0.83\% & 84.15\% & 15.03\% \\
 & Perl & N-Quads & 0.0171 & 0.0005 & 0.0169 & 0.0732 & & 0\% & 100\% & 0\% \\
 & Perl & TriG & 0.0195 & 0.0055 & 0.0007 & 0.0841 & & 0\% & 84.49\% & 15.51\% \\
 & Python & N-Quads & 0.0069 & 0.0011 & 0.0065 & 0.1716 & & 0\% & 100\% & 0\% \\
 & Python & TriX & 0.0063 & 0.0021 & 0.0006 & 0.1325 & & 0.12\% & 84.46\% & 15.42\% \\
\end{tabular}
}
\end{center}
\end{table*}

\subsection{Hash Generation and Checking on Nanopublications}

To test our approach and to evaluate its implementations, we first took a collection of 156,026 nanopublications in TriG format that we had produced in previous work \cite{kuhn2013eswc}. We transformed these nanopublications into the formats N-Quads and TriX using existing off-the-shelf converters. Then, we transformed these into trusty URI nanopublications using the Java implementation. To be able to check not only positive cases (where checking succeeds) but also negative ones (where checking fails), we made copies of the resulting files where we changed a random single byte in each of them (only considering letters and numbers, and never replacing an upper-case letter by its lower-case version or vice versa, as some keywords are not case-sensitive). The resulting six sets of 156,026 files each (three formats, each in two versions: valid and corrupted) were the basis for our evaluation.

The first important result is that all original nanopublications ended up with the same trusty URI, no matter which format was used. This shows that our implementations are successful in handling the content on a more abstract level (i.e. RDF graphs in this case) leading to identical hash values for files that contain the same content but are quite different on the byte level.

Next, we checked the trusty URI of each nanopublication file with the function CheckFile of all implementations that support the respective format. The three right-most columns of Table \ref{tab:checkrdf} show the results. For all valid files (i.e. those we did not corrupt), all implementations correctly verified their trusty URIs. For the corrupted ones, where we randomly changed one byte, the checks almost always failed (by either calculating a different hash value than the one of the trusty URI, or by raising an error that the respective file was not well-formed).

The only corrupted files that were successfully validated were 1,290 TriX files (0.83\%) when running the Java implementation and 181 TriX files (0.12\%) when running the Python implementation. Looking at these concrete cases reveals that they are all harmless. In these cases, the randomly changed byte was not part of the RDF content, but of the meta-information. Due to minor bugs in the used RDF libraries, this meta-information is not sufficiently checked, which leads to accepting the valid content instead of failing because of violated well-formedness.
All our TriX files start with the following two lines:
\begin{ttquote}
<?xml version='1.0' encoding='UTF-8'?>\\
<TriX xmlns='http://www.w3.org/2004/03/trix/tr\\
\verb|  |ix-1/'>
\end{ttquote}
The RDF implementations in Java and Python (or the respective system utilities to load XML files) do not properly check these two lines containing meta-data. Both libraries raise no error if a file starts with something like \texttt{<?Aml} instead of \texttt{<?xml} (106 files); the Python library accepts invalid XML version numbers such as \texttt{1.a} (73 files); and the Java library does not check the TriX namespace argument, raising no error if the argument name is changed to something like \texttt{xmlnZ} (175 files) or the URI is wrong, such as \texttt{.../Prix-1/} (1007 files). In addition, both libraries correctly accept the rare cases (2 files) where the XML version was changed from \texttt{1.0} to \texttt{1.1}, which is the only other valid XML version as of now, though much less common.
All these cases of corrupted files that are successfully verified are harmless because the modified byte has no effect on the internal representation of the RDF content once loaded by the respective library. In a sense, the corrupted byte is automatically corrected in these cases, leaving no trace once the file is loaded.

\subsection{Performance Tests on Nanopublications}

Next, we used the same set of nanopublication files to test the performance of the different modules for checking trusty URI artifacts in different formats. There are two scenarios of how to run such checks: One can run one after the other, as when a small number of nanopublications are manually checked, or one can execute such checks in the form of a batch job in a single program run, which is the preferred procedure to run a large number of checks without supervision. The time required per file is typically much lower in batch mode, as the runtime environment has to start and finalize only once. Therefore it makes sense to have a look at both scenarios.

Table \ref{tab:checkrdf} shows the results of these performance checks for the normal mode (top) and batch mode (bottom).
These results and the ones presented below were obtained on a Linux server (Debian) with 16 Intel Xeon CPUs of 2.27GHz and 24GB of memory.
As expected, the time measurements are much lower in batch mode, but checking is reasonably fast also in normal mode. All average values are below 0.8s (0.03s for batch mode). Using Java in batch mode even requires only 1ms per file. Apart from the runtimes, the two modes had no effect on the results.

\subsection{Performance Tests on Bio2RDF}

\begin{figure*}[t]
\begin{center}
\includegraphics[width=0.75\textwidth]{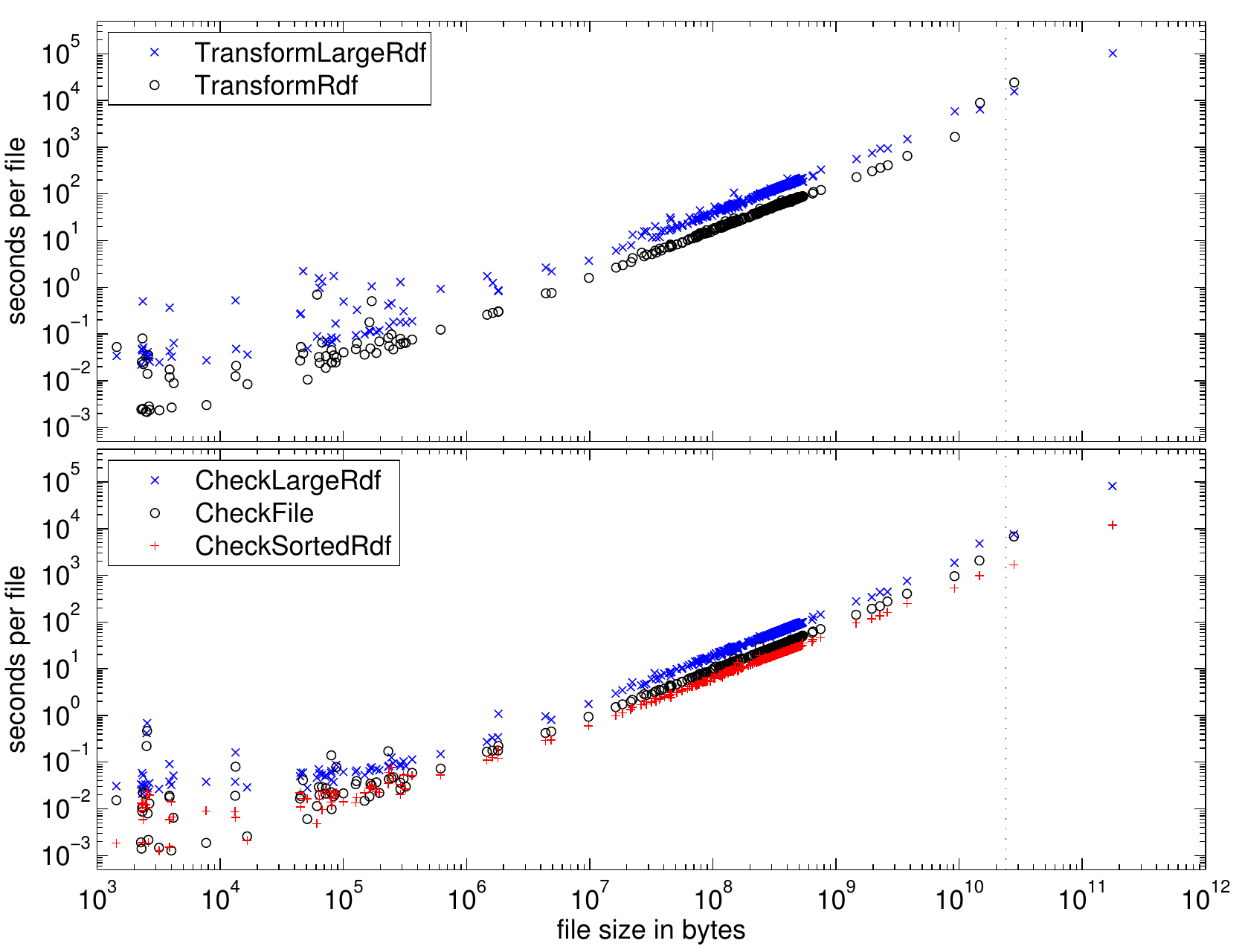}
\caption{Time required for transforming (top) and checking (bottom) files versus file size for the Bio2RDF dataset. The dotted line shows the available memory.
}
\label{fig:bio2rdf-time}
\end{center}
\end{figure*}

The tests above cover only very small RDF files, but our approach should also work for larger files. For that reason, we performed a second evaluation on Bio2RDF ({\small\url{http://bio2rdf.org}}), which is an open-source project focused on the provision of linked data for the life sciences \cite{callahan2013eswc,belleau2008biomedinform}. Bio2RDF scripts convert heterogeneously formatted data (e.g. flat files, tab-delimited files, dataset-specific formats, SQL, and XML) into a common format --- RDF. Over 1 billion triples for 19 resources were made available in the second coordinated release of Bio2RDF \cite{callahan2013eswc}, and mappings to the Semanticscience Integrated Ontology \cite{callahan2013jbiomedsem} have been established.
This second release contains 874 RDF files in N-Triples format, but 16 of them led in our study to well-formedness errors when loaded with the current version of the Sesame library. (These problems might be related to the transition to the new RDF 1.1 standard, and they will be fixed for the next release of Bio2RDF.) This leaves us for the presented study with 858 files of sizes ranging from 1.4kB to 177GB.

Fig. \ref{fig:bio2rdf-time} shows the results of these performance tests. There is a lot of random variation on the lower end, where files are smaller than 10MB and require less than three seconds to be processed. For the upper part, time values nicely follow near-linear trajectories (for the functions that do not load the whole content into memory). When hash calculation involves statement sorting, there is a strict theoretical limit on its performance due to the computational complexity of $O(n \log n)$.
TransformLargeRdf and CheckLargeRdf are superior to their counterparts only for very large files, and CheckSortedRdf is, as expected, faster than the other checking procedures. A large file of 2GB requires about five minutes to be transformed and about two minutes to be checked. Files larger than available memory take more time, but even the largest file of the dataset of 177GB was successfully transformed in 29 hours and checked in about three hours.

\section{Conclusions and Future Work}

We have presented a proposal for unambiguous URI references to make digital artifacts on the (Semantic) Web verifiable, immutable, and permanent. If adopted, it could have a considerable impact on the structure and functioning of the Web, could improve the efficiency and reliability of tools using Web resources, and could become an important technical pillar for the Semantic Web, in particular for scientific data, where provenance and verifiability are important.
Scientific data analyses, for example, might be conducted in the future in a fully reproducible manner within ``data projects'' analogous to today's software projects. The dependencies in the form of datasets could be automatically fetched from the Web, similar to what Apache Maven ({\small\url{http://maven.apache.org}}) does for software projects, but decentralized and verifiable.

As a next step into this direction, we have started to develop a decentralized nanopublication server network \cite{kuhn2014publishing}. Nanopublications are distributed and replicated among such servers and identified by trusty URIs, thereby ensuring that these artifacts remain available even if individual servers are terminated. The current network consists of four servers in four different countries hosting 5 million nanopublications.
In addition, we are working on the concept of \emph{nanopublication indexes} that allow for the definition and identification of small or large sets of nanopublications. Such indexes are nanopublications themselves and, of course, are identified by trusty URIs.

In general, the approach presented in this article might contribute in a significant way to shape the future of publishing on the Web.




\bibliography{trustyurisx}
\bibliographystyle{IEEEtran}

\newpage

\begin{IEEEbiography}[{\includegraphics[width=1in,height=1.25in,clip,keepaspectratio]{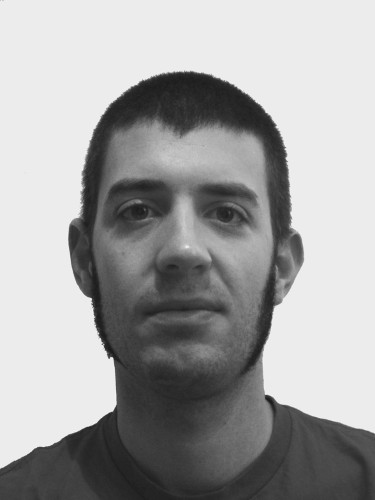}}]{Tobias Kuhn}
is a postdoctoral researcher at ETH Zurich in the Computational Social Science group. His research areas include the Semantic Web, complex systems, user interfaces, computational linguistics, and information systems. He received his PhD at the Institute of Computational Linguistics of the University of Zurich in 2010 for his dissertation on controlled English and knowledge representation. He was also a guest researcher in a software engineering group at the University of Chile, lecturer and researcher at the Intelligent Computer Systems Department of the University of Malta, and postdoctoral associate at Yale University in a bioinformatics lab. Among other topics, he has worked on nanopublications, semantic publishing, controlled natural languages, semantic wikis, citation network analyses, and query interfaces.
\end{IEEEbiography}

\begin{IEEEbiography}[{\includegraphics[width=1in,height=1.25in,clip,keepaspectratio]{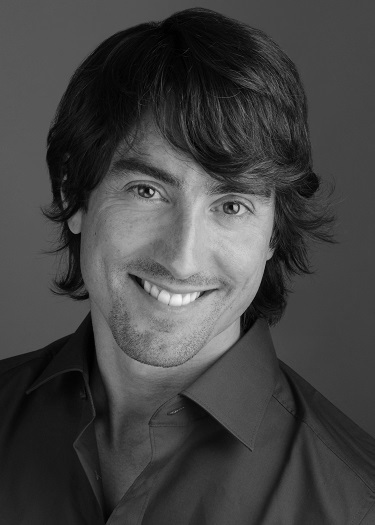}}]{Michel Dumontier}
is an Associate Professor of Medicine (Biomedical
Informatics) at Stanford University. His research focuses on the
development of computational methods to increase our understanding of
how living systems respond to chemical agents. At the core of the
research program is the development and use of Semantic Web
technologies to formally represent and reason about data and services
so as (1) to facilitate the publishing, sharing and discovery of
scientific knowledge, (2) to enable the formulation and evaluation
scientific hypotheses and (3) to create and make available
computational methods to investigate the structure, function and
behaviour of living systems. Dr. Dumontier serves as a co-chair for
the World Wide Web Consortium Semantic Web for Health Care and Life
Sciences Interest Group (W3C HCLSIG) and is the Scientific Director
for Bio2RDF, a widely recognized open-source project to create and
provide linked data for life sciences.
\end{IEEEbiography}


\enlargethispage{-7in}

\end{document}